\documentclass[12pt,preprint]{aastex}

\newcommand{\lsun}{\ensuremath{\mathrm{L}_\odot}}
\newcommand{\kms}{\mbox{km~s$^{-1}$}}
\newcommand{\ammonia}{\mbox{NH$_3$}}
\newcommand{\formamide}{\mbox{NH$_2$CHO}}
\newcommand{\methanol}{\mbox{CH$_3$OH}}
\newcommand{\methanolap}{\mbox{CH$_3$OH~A$^+$}}

\newcommand{\methylcyanide}{\mbox{CH$_3$CN}}
\newcommand{\dimethylether}{\mbox{CH$_3$OCH$_3$}}
\newcommand{\hcthreen}{\mbox{HC$_3$N}}
\newcommand{\methylformate}{\mbox{HCOOCH$_3$}}
\newcommand{\ethylcyanide}{\mbox{CH$_3$CH$_2$CN}}
\newcommand{\HII}{H {\sc ii}}
\newcommand{\cmcub}{\mbox{cm$^{-3}$}}
\newcommand{\cmsq}{\mbox{cm$^{-2}$}}
\newcommand{\nhtwo}{\mbox{$N_{\rm H_2}$}}

\newcommand{\CeiO}{C$^{18}$O}
\newcommand{\hcoplus}{HCO$^+$}
\newcommand{\msun}{\ensuremath{\mathrm{M}_\odot}}

\shorttitle{Millimeter Spectroscopy of G34.26+0.15}
\shortauthors{Mookerjea et al.}

\begin{document}
\title{Kinematics and Chemistry of the Hot Molecular Core in G34.26+0.15 at High
Resolution}  
\author{B. Mookerjea, E. Casper, L. G. Mundy}
\affil{Department of Astronomy, University of Maryland, College Park,
MD 20742}
\author{L. W. Looney}
\affil{Astronomy Department, University of Illinois, 1002 W Green St,
Urbana, IL 61801} 

\begin{abstract}
We present high angular resolution ($\sim 1$\arcsec) multi-tracer
spectral line observations toward the hot core associated with
G34.26+0.15 between 87--109~GHz. We have mapped emission from (i)
complex nitrogen- and oxygen-rich molecules like \methanol, \hcthreen,
\ethylcyanide, \formamide, \dimethylether, \methylformate; (ii)
sulfur-bearing molecules like OCS, SO and SO$_2$; and (iii) the
recombination line H53$\beta$. 

The high angular resolution enables us to directly probe the hot
molecular core associated with G34.26+0.15 at spatial scales of
0.018~pc. At this resolution we find no evidence for the hot core being
internally heated. The continuum peak detected at $\lambda=2.8$~mm is
consistent with the free-free emission from component C of the
ultracompact \HII\ region. Velocity structure and morphology outlined by
the different tracers suggest that the hot core is primarily energized
by component C. Emission from the N- and O-bearing molecules peak at
different positions within the innermost regions of the core; none are
coincident with the continuum peak.  Lack of high resolution
complementary datasets makes it difficult to understand whether the
different peaks correspond to separate hot cores, which are not resolved
by the present data, or manifestations of the temperature and density
structure within a single core.

Based on the brightness temperatures of optically thick lines in our
sample, we estimate the kinetic temperature of the inner regions of the
HMC to be  $160\pm30$~K.  Comparison of the observed abundances of the
different species in G34.26+0.15 with existing models does not produce a
consistent picture. There are uncertainties due to: (i) the
unavailability of  temperature and density distribution of the mapped
region within the hot core, (ii) typical assumption of centrally peaked
temperature distribution for a hot core with an accreting protostar at
the center, by the chemical models, an aspect not applicable to
externally heated hot cores like G34.26+0.15 and (iii) inadequate
knowledge about the formation mechanism of many of the complex
molecules.

\end{abstract}
\keywords{Stars:formation -- ISM:abundances--ISM:individual
(G34.26+0.15) -- ISM:molecules -- radio lines: ISM -- radio
continuum:ISM}

\section{Introduction}

Hot cores are compact (0.1~pc), warm ($\sim$ 100--300~K), dense
(10$^7$ H nuclei~\cmcub) clouds of gas and dust near or around sites
of recent star formation \citep[see
e.g.][]{kurtz2000,cesaroni2005,vandertak2005}. The hot-core phase is
thought to last about 10$^5$~yr \citep{vandishoeck1998} to 10$^6$~yr
\citep{garrod2006} and represents the most chemically rich phase of
the interstellar medium, characterized by complex molecules like
\methanol, \methylcyanide, \methylformate, \dimethylether\ and
\ethylcyanide. The complex chemical and physical processes occurring
in the hot-cores are not fully understood. Until recently hot cores
were thought to be associated with high-mass protostars (M$\geq$
8~\msun) only and to represent an important phase in their evolution
toward ultracompact and compact \HII\ regions.   The central
energizing source for a number of hot cores have been identified using
high angular resolution mid-infrared (MIR) and millimeter continuum
observations \citep[e.g.][]{debuizer2003,beltran2004}. These
detections strengthen the idea of hot molecular cores (HMCs) as
representing a stage in the evolutionary sequence of massive
protostars. However, there are non-negligible examples of hot cores
that are in the vicinities of UC \HII\ regions and appear to be only
externally heated.  For these sources it may well be argued that the
hot cores can also arise as chemical manifestations of the effect of
the UC \HII\ regions on their environments, rather than being only the
precursors of UC \HII\ regions.

The chemical models are still far from providing a unique
interpretation of the hot core chemistry, and would benefit from high
angular resolution continuum and spectroscopic observations suitable
to understand the temperature and density distributions of the cores.
In particular, the strong sensitivity of surface and gas-phase
chemistry to dust temperature and gas density highlights the
importance of the study of the abundances of complex molecules in
different regions with varied physical characteristics. Furthermore,
all the existing chemical models assume the hot cores to be internally
heated and have radially varying density and temperature profiles
\citep{millar1997,nomura2004,garrod2006,doty2006}; the models do not yet
provide a consistent treatment of the externally heated hot cores.

We present high angular resolution (1\arcsec) interferometric
observations with the Berkeley-Illinois-Maryland Association
(BIMA\footnote{The BIMA array was operated by the Berkeley-Illinois
Maryland Association under funding from the National Science
Foundation. BIMA has since combined with the Owens Valley Radio
Observatory millimeter interferometer, moved to a higher site, and was
recommissioned as the Combined Array for Research in Millimeter
Astronomy (CARMA) in 2006}) array of the well-studied hot core
associated with the UC \HII\ region G34.26+0.15 located at a distance
of 3.7~kpc \citep{kuchar1994}. The \HII\ region is a prototypical
example of cometary morphology, which may be due to the bow-shock
interaction between an ambient molecular cloud and the wind from an
energetic young star moving supersonically through the cloud
\citep{wood1989,vanburen1990}.  \ammonia\ observations with the VLA
\citep{heaton1989} show that the highly compact molecular cloud
appears to be wrapped around the head of the cometary \HII\ structure,
with the ionization front advancing into the cloud. 

G34.26+0.15 has been extensively studied in radio continuum
\citep{turner1974,reid1985,wood1989} and radio recombination lines
\citep{garay1985,garay1986,gaume1994,sewilo2004}. At radio continuum
frequencies, it exhibits several components: two UC \HII\ regions called
A \& B, a more evolved \HII\ region with a cometary shape named
component C, and an extended ring-like \HII\ region with a diameter of
1\arcmin\ called component D \citep{reid1985}.  Molecular gas has been
mapped in \ammonia, \hcoplus, SO, \methylcyanide\ and CO
\citep{henkel1987,heaton1989,carral1992,heaton1993,akeson1996,watt1999}.

The hot core associated with G34.26+0.16 has been the target of
chemical surveys using single-dish telescopes
\citep{macdonald1996,hatchell1998} in which complex molecules
characteristic of hot cores were detected.  Molecular line
observations suggest that the hot core does not coincide with the
\HII\ region component C; it is offset to the east by at least
2\arcsec\ {and shows no sign of  being internally heated}
\citep{heaton1989,macdonald1995,watt1999}.  Based on narrow-band
mid-infrared  imaging of the complex, \citet{campbell2000} concluded
that the same star is responsible for the ionization of the \HII\
component C and heating the dust but is not interacting with the hot
core seen in molecular emission. At a resolution of 12\arcsec,
\citet{hunter1998} also found the peak of the 350~\micron\ emission to
be coincident with the  component C of the UC \HII\ region. 

In this paper we use the BIMA observations to study the energetics,
chemistry and kinematics of the molecular gas contributing to the hot
core emission associated with G34.26+0.15.

\section{Observations}

Observations of the source G34.26+0.15 were acquired  with the
ten-element BIMA interferometer between 1999 December and March 2000
at three frequency bands centered approximately at 87, 107 and 109~GHz
using the A \& B configurations of the array.  Due to technical
difficulties only nine antennas could be used for the observations at
87~GHz. Table~\ref{obsdetails} presents a log of the observations,
including the typical system temperatures in the different
configurations  presented here.  The primary FWHM of the
array is between 132\arcsec\ and 106\arcsec\ at frequencies between 87
and 109~GHz.  The correlator was configured to split each frequency
into four windows per sideband, two of which had bandwidths of 100~MHz and 64
channels each and the remaining two had bandwidths of 50~MHz and 128
channels each.

The sources Uranus, 1830+063 and 1771+096 were observed as the
primary flux calibrator, the phase calibrator and the secondary
calibrator, respectively. However owing to the consistently poor
quality of 1830+063 observations and the sparsity of Uranus observations,
we have used 1771+096 as both the phase and primary flux calibrator.
The flux of 1771+096 was determined from each of the six datasets,
using the MIRIAD task bootflux with Uranus as the primary calibrator.
The average final flux for 1771+096 is 2.3~Jy; we estimate the
absolute flux calibration error to be 10\%.  The pointing and phase
center used for  mapping the region around G34.26+0.15 is
$\alpha_{2000}$ = 18$^h$53$^m$18\fs55 $\delta_{2000}$ = 1\arcdeg
14\arcmin 58\farcs2 and the $V_{\rm lsr}$ = 58~\kms.

The data were reduced using the MIRIAD \citep{sault1995} software
package. Continuum maps were constructed by averaging over spectral
windows that did not contain line emission. In order to achieve better
phases the continuum maps were iteratively self calibrated. 
The spectral line maps within a particular spectral window were made
by first fitting a low-order polynomial to the visibilities in the
channels not contaminated by any line emission and then subtracting
the fit to the continuum from the individual channels.

\section{Results}

\subsection{Spectral Lines}

The observed spectral lines were identified using the spectral line
catalog by \citet{lovas1979}.  We have detected multiple transitions of
methanol (\methanol), two transitions of ethyl cyanide (\ethylcyanide)
and single transitions of methyl formate (\methylformate), dimethyl
ether (\dimethylether), formamide (\formamide), \hcthreen, OCS (and its
$^{34}$S, $^{13}$C isotopomers), SO, SO$_2$ and the hydrogen
recombination line H53$\beta$.  Table~\ref{speclines} presents
spectroscopic details of the molecular lines that were detected in the
present data set.

Table~\ref{basic_results} summarizes the basic results of the BIMA
observations:  synthesized beam sizes, the peak intensity, the peak
brightness temperature, the central velocity ($v_{\rm cen}$), the
velocity width ($v_{\rm fwhm}$)  and the achieved rms per channel of
the interferometric maps, for each of the observed spectral lines.
The $v_{\rm cen}$ and $v_{\rm fwhm}$ were determined from Gaussian
fits to the line profiles. In addition to the 10\% error in absolute
calibration we estimate the statistical errors in intensities
to be $\sim 10$\%.

\subsection{Continuum at $\lambda=2.8$~mm}
\label{continuum}

\begin{figure}
\begin{center}
\includegraphics[angle=0,width=7.5cm]{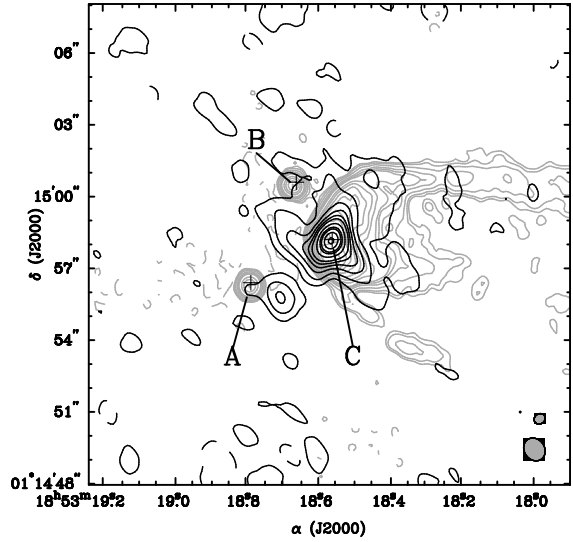}
\caption{Overlay of the contours (black) of the 2.8~mm continuum
emission detected with BIMA with the contours (grey) of the 2~cm
free-free continuum emission observed by \citet{sewilo2004}.  For the
$\lambda=2.8$~mm continuum image the contour levels are  -3, 3, 10,
20, 30, 40, 50, 70, 90, 100, 120, 140 of the noise level of
20~mJy~beam$^{-1}$.  For the $\lambda=2$~cm continuum image the
contour levels are at -3, 3, 5, 10, 20, 30, 40, 80, 120, 180, 250,
350, 450, 600, 800, 850 times the noise level of 0.3~mJy~beam$^{-1}$.
The synthesized beams are shown on the right-hand corner of the plot.
The synthesized beam at $\lambda=2.8$~mm is 1\farcs00$\times$0\farcs85
with P.A. = 35\arcdeg\ and the beam at $\lambda=2$~cm is
0\farcs48$\times$0\farcs41 with P.A.=21\arcdeg.
\label{cont_cont}}
\end{center}
\end{figure}

Figure~\ref{cont_cont} shows an overlay of the $\lambda=$2.8~mm
continuum image with the $\lambda=$2~cm continuum emission observed by
\citet{sewilo2004}.  The 2.8~mm continuum is peaked at the
nominal map center ($\alpha_{2000}$ = 18$^h$53$^m$18\fs55,
$\delta_{2000}$ = 1\arcdeg 14\arcmin 58\farcs2) and coincides with the
cometary component C of the G34.26+0.15 UC \HII\ region complex
\citep{heaton1989,gaume1994,sewilo2004}. As will be discussed in
Sec.~\ref{intmap_peak} the continuum peak detected at 2.8~mm 
does not coincide with the emission peaks of the lines originating in
the hot molecular gas. Although continuum at 2.8~mm shows a single
peak, most of the radio continuum maps at cm wavelengths show the
component C to consist of two emission peaks C1 and C2, separated
by $\sim 1\farcs4$. However \citet{sewilo2004} suggest that C1 and C2
are not two separate continuum components, but they correspond to two
regions of maximum emission in the nebula. Recently
\citet{avalos2006} mapped the G34.26+0.15 region at 43~GHz using VLA
and they find a single peak similar to the millimeter maps.

Based on the radio continuum spectrum the flux density due to the
free-free emission from the components A \& B is expected to be $\sim
100$~mJy at 100~GHz \citep{avalos2006}.  This is consistent with our
observations where we do detect some emission from the UC components A
and B, though these components are not resolved in our maps.  At 107~GHz
($\lambda=2.8$~mm) we measure a peak flux density of 2.9 Jy/beam and an
integrated intensity of 6.7$\pm0.4$~Jy estimated by fitting a Gaussian
of 1\farcs6$\times$1\farcs4 to the central source. In addition to the
given statistical error for the continuum flux density, there is a
0.7~Jy error corresponding to the uncertainty in the absolute flux
calibration.  The 2.8~mm continuum flux density measured here is
consistent with previous observations at similar resolutions
\citep{akeson1996}.  \citet{heaton1989} measured an integrated radio
continuum flux of 5~Jy at 1.3~cm (24~GHz) for the component C, which,
when combined with the 2.8~mm flux indicates a spectral index of 0.2.
This spectral index is significantly flatter than the index of 0.45
derived by \citet{watt1999} for the combined emission of all 3
components of the UC \HII\ region. At a resolution of 1\arcsec, the
2.8~mm emission is clearly dominated by the free-free emission from
component C of the UC \HII\ region and does not arise from an embedded
source that could be energizing the HMC internally. Dust emission may be
responsible for the eastern extension of the emission at a 10\% level.  

\subsection{Integrated Intensity Maps}
\label{intmap_peak}

\begin{figure}
\vspace*{-25mm}
\begin{center}
\includegraphics[angle=0,width=15.0cm]{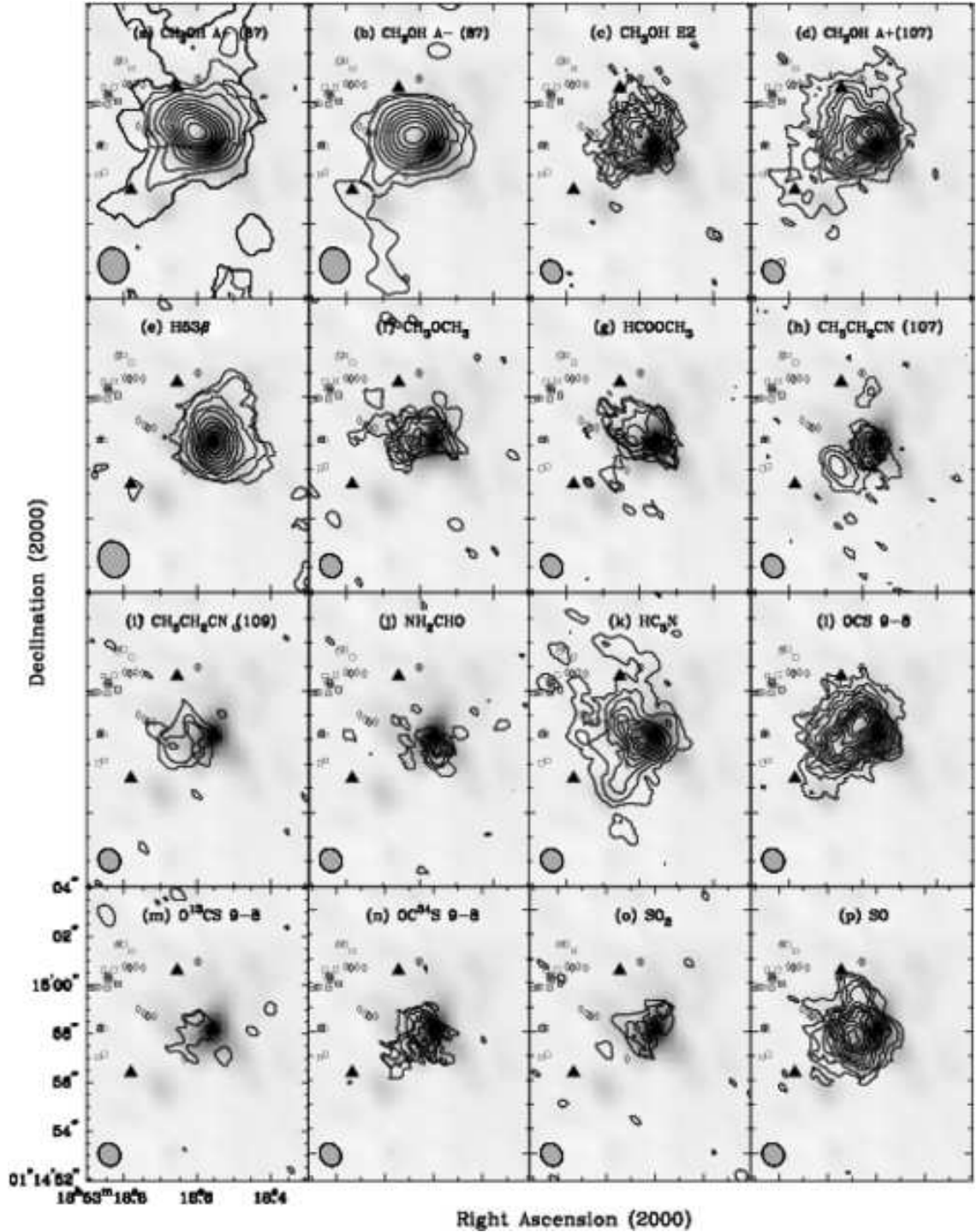}
\caption{Integrated intensity maps of the spectral lines detected in
G34.26+0.15. The species are marked within individual panel. In case
of multiple transitions of the same species, the nominal frequency
band is also noted. Details of the lines are presented in
Table~\ref{speclines}. The noise levels (in Jy~beam$^{-1}$ \kms) are :
{\em (a)} 0.58, {\em (b)} 0.40, {\em (c)} 0.26, {\em (d)} 0.45, {\em
(e)} 0.61, {\em (f)} 0.58, {\em (g)} 0.30, {\em (h)} 0.48, {\em (i)}
30, {\em (j)} 0.24, {\em (k)} 0.50, {\em (l)} 0.35, {\em (m)} 0.58,
{\em (n)} 0.22, {\em (o)} 0.46 and {\em (p)} 0.32. The contour levels
in panels {\em (a), (f), (g), (h), (i), (j), (m), (n)} and {\em (o)}
are in steps of the noise level, starting from twice the noise level.
The contour levels in the remaining panels are in steps of twice the
noise level, starting from twice the noise level.  The negative
contours are not shown to simplify the image. The synthesized beams
are shown on the left hand corner of each panel. The filled $\Delta$
mark the positions the components A, B, and C \citep{heaton1989} of
the UC \HII\ region G34.3+0.2. The squares and diamonds signify
positions of H$_2$O and OH masers respectively \citep{forster1989}.
\label{allmap}}
\end{center}
\end{figure}

Figure~\ref{allmap} shows the integrated intensity maps of the
different spectral line emission detected in the region around
G34.26+0.15 overlaid with the continuum emission at $\lambda=2.8$~mm
and  the positions of the H$_2$O and OH masers detected in the
region \citep{forster1989}. The emission from none of the molecular
species coincides with the positions of the H$_2$O and OH masers.

\begin{figure}
\begin{center}
\includegraphics[angle=0,width=10.0cm]{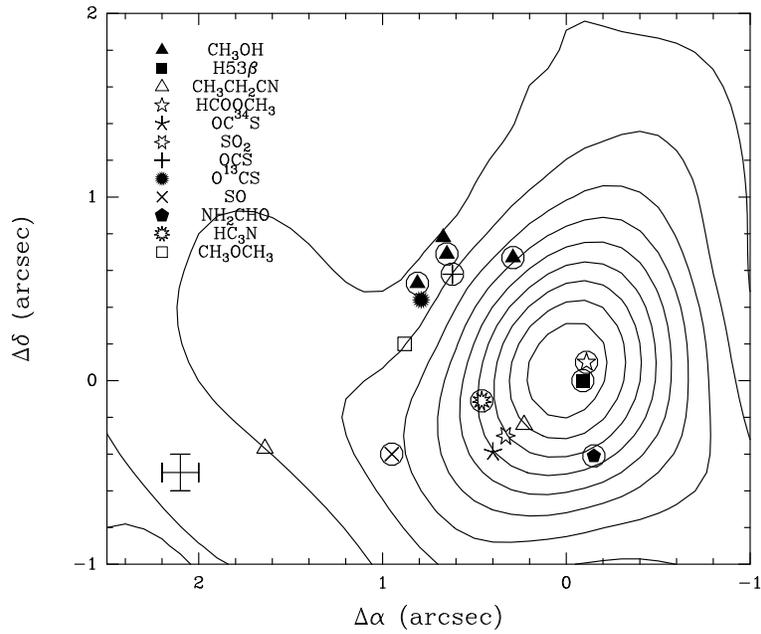}
\caption{Peak positions for the different spectral lines detected in
G34.26+0.15 overlaid with the $\lambda=2.8$~mm.
The symbols with additional circles around them denote that these 
correspond to single-peaked intensity distributions. The errorbars on
the left hand bottom corner show the positional
uncertainty.
\label{peaks}}
\end{center}
\end{figure}

At a resolution of 1\arcsec\ the integrated intensity distribution for
the different chemical species primarily show three kinds of
morphological structures: {\em single-peaked, double-peaked} and {\em
irregular}.  Figure~\ref{peaks} shows the positions of the absolute
maxima of the integrated intensities for the different species
overlaid with the $\lambda=2.8$~mm continuum map of G34.26+0.15.
The absolute positional accuracy is estimated to be 0\farcs1.  We note
that it is possible to identify primarily two regions where the maxima
due to the transitions of the different species are localized: the
first lying slightly to the south-east of the continuum peak and the
second lies to the north-east of the 2.8~mm continuum peak.

{\em (a) Single-Peaked : } The spectral lines of \methanol\ (with the
exception of the $15_{-2}$--15$_{1}$ E2 transition), \hcthreen,
NH$_2$CHO, SO and OCS, as well as the H53$\beta$ recombination line
show well-defined single peaked structures. \methylformate\ also shows
a single-peaked structure though it is somewhat poorly defined as
compared to the rest. Of these spectral lines only the peaks of the
H53$\beta$ and the \methylformate\ emission coincide with the peak of
the continuum at $\lambda=2.8$~mm, while the peaks due to \methanol\
and OCS occur to the north-east of the continuum peak. The peak due to
SO lies 1\arcsec\ to the south-east of the 2.8~mm continuum peak.  The
peaks of \hcthreen\ and NH$_2$CHO lie close enough ($\sim$ 0\farcs4
south-east) to but not exactly at the position of the 2.8~mm continuum
peak. At a first glance, it appears that the more abundant species with
higher optical depths tend to show a single-peaked structure. We
revisit the question of optical thickness of the lines in
Sec.~\ref{kintemp}.

{\em (b) Double-Peaked :} The spectral lines of CH$_3$CH$_2$CN and
\dimethylether\ show rather well-defined double-peaked structure.
Figure~\ref{peaks} suggests that the peaks corresponding to the two
transitions of \ethylcyanide\ occur at rather different positions. In
fact, they correspond to the two peaks that appear in both
transitions, with essentially reversed ratios of relative intensities.
The double-peaks of \dimethylether\ are also not aligned with those of
\ethylcyanide, as \dimethylether\ emission as a whole is somewhat
shifted to the north of the continuum peak and extends more along the 
east-west direction in contrast to the \ethylcyanide\ emission.

{\em (c) Irregular Shaped:} The integrated intensity distribution of
O$^{13}$CS, OC$^{34}$S and SO$_2$ appear to be irregularly shaped and
do not have enough signal-to-noise in order to show well-defined
peaks. The positions shown in Figure~\ref{peaks} correspond to the
absolute maxima within the emitting region and represent the
approximate location of the enhancement in emission. These should not
be taken too literally as peaks having significance comparable to the
peaks for the single-peaked species. 

Figure~\ref{peaks} shows that almost all peak positions to the
north-east of the continuum peak correspond to single-peaked emission
structure. Among the tracers peaking to the south-east, only
\hcthreen\  and SO  are single-peaked and \ethylcyanide\ is
double-peaked.  The remaining tracers have irregular intensity
distribution with absolute maxima at the indicated positions.  

These two primary locations of the peaks of the spectral lines,
(Fig.~\ref{peaks}) may either correspond to different emitting clumps or they
could be  manifestation of the differences in the temperature and
density that leads to the variation of the chemical abundances within
the same clump. We revisit both these concepts in
Sec.~\ref{discussion}. In Sec.~\ref{chemabund} we derive the chemical
abundances of the different chemical species at the two nominal peak
positions identified by the \methanol\ $7_2$--$6_3$A$^+$ peak to the
north-east and the \hcthreen\ 12--11 peak to the south-east of the
continuum peak.

\subsection{Velocity Gradients}

\begin{figure}
\begin{center}
\includegraphics[angle=0,width=12.0cm]{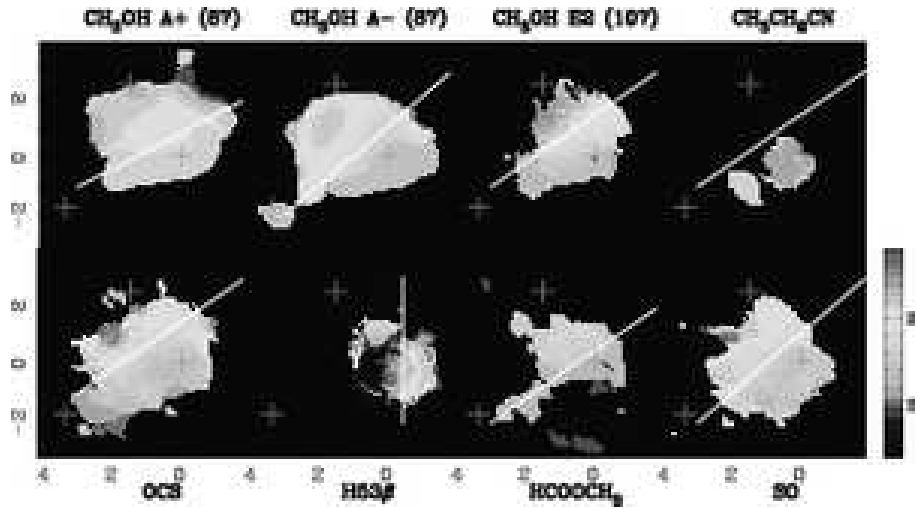}
\caption{Centroid velocity maps of selected species observed in
G34.26+0.15.  The colorscale extending between 52 and 64~\kms\ is
shown to the right of the figure. The drawn straight lines show the
direction perpendicular to the velocity gradients approximately.  The
`+' mark the positions the components A, B, and C \citep{heaton1989}
of the UC \HII\ region G34.3+0.2.
\label{velmap}}
\end{center}
\end{figure}

Figure~\ref{velmap} shows the centroid velocity distribution of some
of the selected species for which the signal to noise was sufficient
over large areas to generate the maps. With the exception of the
hydrogen recombination line and  the limited sensitivity
\ethylcyanide\ line , the center velocity of all molecular lines
changes from 56~\kms\ in the south-west to 62~\kms\ in the north-east
over a spatial scale of 0.08~pc in a direction approximately with a
position angle of 40$\pm10$\arcdeg. The direction and magnitude of the
velocity gradient match well with other measurements of molecular gas
velocity gradients towards G34.26+0.15.  Using observations at angular
resolutions similar to the present dataset \citet{heaton1989} derived
a velocity gradient of $\geq 50$~\kms\ pc$^{-1}$, while
\citet{carral1992} and \citet{heaton1993} observed a velocity gradient
of 6$\pm1$~\kms\ on 0.3 to 0.75~pc scales. \citet{akeson1996} derived
a velocity gradient of 15$\pm5$~\kms\ on 0.2 to 0.3~pc scales.
\citet{watt1999} observed similar velocity gradients as we find, using
only \methylcyanide.  SiO emission primarily tracing outflow activity
was not detected at the position of the hot core, it rather appears to
surround the hot core \citep{hatchell2001}. This implies that shocks
do not play an important role in regulating the hot core chemistry in
G34.26+0.15.

The ionized and molecular gas in G34.26+0.15 show rather different
velocity gradients, both in direction and magnitude.  The hot
molecular gas consistently shows this SW-NE velocity gradient, while
the ionized gas has a total gradient of $\approx 35$~\kms\ in a
direction perpendicular to the symmetry axis of the cometary UC \HII\
region \citep{gaume1994}. A smaller velocity gradient is detected in
the ionized gas parallel to the symmetry axis, {\em i.e.} along the
east-west direction. In addition, the ionized gas shows typical
velocity widths of $\sim 50$~\kms\
\citep{garay1986,gaume1994,sewilo2004}.  The hydrogen recombination
line, H53$\beta$ shows an emission peak exactly coincident with the
continuum peak at $\lambda=2.8$~mm and a velocity gradient similar to
the minor component of the velocity gradient as identified in the
ionized gas. Similar to all the other recombination lines detected in
the region, the H53$\beta$ line has a linewidth much
larger than all the other molecular species. These large widths of the
recombination lines show conclusively that the ionized gas is driven
primarily by the \HII\ region dynamics.

Comparison of Figs.~\ref{allmap} and \ref{velmap} show that the
observed velocity gradient in the hot molecular gas is along the minor
axis of the source emission for all species. This is also consistent
with the results of \citet{watt1999}. For a core with significant
rotation, the velocity gradient is expected to be along the major axis
of the source. This argues strongly against the velocity gradients
arising due to the gravitationally bound rotation of a circumstellar
slab or disk as was proposed by \citet{garay1986}.  Several authors
have interpreted the observed velocity distribution of the UC \HII\
region G34.26+0.2C in terms of mainly two models: the moving star bow
shock model and the champagne outflow model. However,
\citet{gaume1994} point out that none of these models satisfactorily
explain the observed velocities in the region. These authors proposed
a model in which the velocities in the region are governed by the
stellar winds from the components A and B. Further, the ionized gas
which is photoevaporated directly from the hot, dense UC molecular
core by the exciting star of the cometary \HII\ region G34.26+0.2C,
flows from the molecular core toward a region of lower density to the
west. In this model, it is further proposed that the molecular core
and the components A and B are at a slightly larger distance from the
Earth than G34.3+0.2C. The velocity pattern derived from the various
species observed here is consistent with the ``wind and
photoevaporation" model proposed by \citet{gaume1994}, and this
further strengthens the conclusions of \citet{watt1999}, which were
derived based only on the velocity patterns of  \methylcyanide.

\section{Energetics of the hot core in G34.26+0.15}

Hot cores are typically proposed to be precursors of high mass stars
\citep[e.g.][]{cesaroni2005}. The center of the hot core is identified
with a collapsing, and rapidly accreting high mass protostar.  High
angular resolution mid-infrared (MIR) observations have successfully
detected the energizing sources for several  hot cores like
G11.94-0.62, G45.07+0.13, G29.96-0.2 etc.
\citep{debuizer2002,debuizer2003}. Using sub-arcsecond resolution
centimeter and millimetre molecular line and continuum observations
the existence of multiple deeply embedded UC and hypercompact \HII\
regions contributing to the formation of the HMCs G10.47+0.03 and
G31.41+0.31 have been substantiated \citep{beltran2004,beltran2005}.
Most of the HMCs thus represent a stage in the evolutionary sequence
of massive protostars.  

However, in the case of the hot molecular gas in G34.26+0.15 the existing
continuum observations in mid-infrared, far-infrared, sub-millimeter
and millimeter have consistently shown the peak of the dust emission
to be coincident with the UC \HII\ region component C and not at the
position of the hot core \citep{hunter1998,campbell2000,campbell2004}.
In particular, using MIR observations with 1--2\arcsec\ resolution
\citep{debuizer2003} have reported non-detection of any MIR source
associated with the HMC  in G34.26+0.15.  Thus
G34.26+0.15 similar to the Orion Compact Ridge and W3(OH)
\citep{wyrowski1999} represents the alternative scenario for hot
cores, which are externally heated and viewed upon as manifestation of
gas shocked and heated by the expanding ionization front and the
stellar winds arising from the \HII\ regions.  Medium angular
resolution MIR and FIR observations further suggest that the HMC
is isolated from the component C of the \HII\ region and is heated by
stellar photons from A and B components and not shocks
\citep{campbell2000,campbell2004}.  \citet{watt1999} compared the
\methylcyanide\  and \CeiO\ peaks  with the \ammonia\ peaks seen in
arcsecond resolution observations by \citet{heaton1989} and found that
none of the peaks coincide with either component C of the \HII\ region
or with each other. These suggest that the hot molecular gas detected
in G34.26+0.15 traces a layer of a core that is being externally
heated by shocks and stellar photons. However, non detection of SiO
emission from the position of the hot core rules out any significant
role played by shocks in determining the hot core chemistry
\citep{hatchell2001}.

The arcsecond resolution continuum and molecular line observations
presented in this paper provide further evidence to this proposed model
of the hot core in G34.26+0.15 being externally heated.  The continuum
map at 2.8~mm shows a single peak coincident with the radio continuum
peak, and it is offset from the peaks of all molecular lines. The
molecular line maps extending over a region of
9\arcsec$\times$12\arcsec\ ($\alpha\times\delta$) show that (i) the
emission from different species are not centrally peaked, (ii) the
molecular peaks are not co-spatial, (iii) none of the molecular line
emission peaks at the continuum peak and (iv) none of the molecular line
peaks coincide with the H$_2$O and OH masers, tracing locations of
shocks due to the propagation of the ionization front from the UC \HII\
regions,  detected in the region \citep{forster1989}.

The relative offsets between the peaks in the different molecular line
emission further brings out the possibility of there being multiple
HMCs within the beam.  We note that the NE and the SE peaks detected
in the G34.26+0.15 region are separated along the north-south
direction by 0\farcs8, which at a distance of 3.7~kpc translates to
0.014~pc.  For comparison, the Hot Core and the Compact Ridge regions
in the Orion-KL cloud  at a distance of 450~pc are separated by
0.018~pc. We consider the angular resolution of the present
observations to be insufficient to either clearly resolve or rule out
the possibility of the two peaks being two different cores.  These
results differ substantially from the single-dish observations which
have so far suggested that the emission from the different molecular
species originate from within a single bound core.  

The observations of G34.26+0.15 presented here, with the highest
angular resolution to date, suggest multiple externally heated HMCs, a
hypothesis which is a matter of further study.  In the absence of
observations of multiple transitions of molecular species at 1\arcsec\
resolution, it is not possible to derive the density and temperature
distributions  appropriate to decipher factors
contributing to the relative offsets of the different emission peaks.
Availability of such temperature and density profiles is also crucial
for deriving accurate estimates for the abundance profiles for the
detected complex molecular species.

\section{Available chemical models for hot cores and G34.26+0.15}

Existing and recent high angular resolution observations of the HMC
associated with G34.26+0.15 strongly favour that the source is
externally heated. However there are no observations available that
constrain (i) the geometry of the hot core material with respect
to the external heating source; (ii) the density structure of the core
material, {\em e.g.} whether it shows a simple gradient in a particular
direction or it is clumpy or it is a uniform density swept up region;
or (iii) the geometry and multiplicity of the core along the line of
sight. We propose the present dataset as a building block for future
observations which would help clarify some of the outstanding issues.

The rather ``unconventional" and complicated geometry of the HMC
associated with G34.26+0.15 poses further challenge to deriving a
consistent chemical model for the source. All existing chemical models
for HMCs, including those constructed specifically for G34.26+0.15
\citep[e.g.][]{millar1997,nomura2004} explicitly consider a centrally
energized spherical cloud forming a massive protostar in two stages
involving collapse to high densities ($\sim 10^7$~\cmcub) followed by a
warm-up phase resulting in the rich chemistry of the hot cores. The
models for the HMC in G34.26+0.15 are primarily based on single dish
observations and assume that the source consists of a hot ultracompact
core (UCC) with a radius less than 0.025~pc and $n(H_2) = 2\times
10^7$~\cmcub, a compact core (CC) with a radius of about 0.1~pc and a
density of 10$^6$~\cmcub, surrounded by a massive halo extending out to
3.5~pc with a H$_2$ density that falls off as r$^{-2}$ \citep[e.g.][and
the references therein]{millar1997}. The latest chemical model for
G34.26+0.15 by \citet{nomura2004} is more sophisticated in terms of
using the density and temperature profiles derived from the dust
continuum observations. These density and temperature profiles are
subsequently used as inputs for the chemical model. However, the density
and temperature distributions are primarily constrained by single-dish
continuum observations and the model considers a centrally peaked
spherically symmetric structure that is isothermal up to a radius of
0.05~pc ($\sim 3$\arcsec). Further it assumes the core of the region to
be collapsing to form a massive protostar. Most of these assumptions do
not conform to the results of high angular resolution molecular line
observations presented in this paper and also by \citet{watt1999} and
\citet{heaton1989}.  Similar arguments may be invoked to show the
inappropriateness of the radiative transfer models developed by
\citet{hatchell2003} for the HMC in G34.26+0.15.

\section{Estimate of Kinetic Temperature
\label{kintemp}}

In the absence of high resolution multi-line spectroscopic observations
suitable to derive the temperature and density distributions, we have
adopted a simplistic approach to derive an estimate of the kinetic
temperature from the observed brightness temperatures.

Using the brightness temperatures of OC$^{34}$S 9--8 (50~K) and OCS
9--8 (96~K) at the position of the intensity peak of OC$^{34}$CS
(Table~\ref{basic_results}),  and the relative abundance of the
isotopomers $^{32}S$/$^{34}S$ $\sim 22$ \citep{wilson1994} we derive
$\tau_{34} = 0.7$ and $\tau_{32}=15.4$.  Hence OCS 9--8 is optically
thick and its peak brightness temperature (167~K) is a reasonably good
estimate of the gas kinetic temperature.  We further note that the
peak brightness temperature of \methanol\ 3$_1$--4$_0$ A$^+$, the
brightest and the most optically thick of the methanol lines observed
here is 187~K: the peak brightness temperatures of SO and \hcthreen\
are  147~K and 143~K, respectively.  Thus beyond the calculated high
opacity values of OCS 9--8, the rather high peak brightness
temperatures of \methanol\ 3$_1$--4$_0$ A$^+$, SO and \hcthreen\ suggest
that all these lines are optically thick and their peak brightness
temperatures provide a firm lower limit to the true kinetic
temperature of the UCC of G34.26+0.15.  Based on these, we estimate the
kinetic temperature to be 160$\pm30$~K and use 160~K for the
subsequent analysis.

Most of the available estimates of kinetic temperature of the hot core
region G34.26+0.15 are rotation temperatures derived using a variety
of molecules. Based on an LVG analysis of CH$_3$CN \citep{watt1999}
derived a gas temperature between 80--175~K, while using the K-ladder
of the same molecule \citet{akeson1996} estimated a kinetic
temperature of $250\pm100$~K.  \citet{millar1995} derived a rotation
temperature of 125~K from C$_2$H$_5$OH observations, while
\citet{henkel1987} derived a rotation temperature of $225\pm75$~K from
ammonia observations. 

The errorbars in the kinetic temperatures available from literature
are rather large and they are all consistent with our derived value of
160$\pm30$~K. The higher angular resolution of our observations better
probe the small clumps in contrast to the other observations.  We note
that the molecules of which  rotational transitions  were used to
derive the kinetic temperatures have similar dipole moments. However,
comparison with temperatures derived from single dish observations
with the present results is uncertain due to multiple issues: beam
dilution, contamination from large scale emission, etc.

\section{Column Densities of different species
\label{chemabund}}

We have derived the column densities of the different observed species
assuming  Local Thermal Equilibrium (LTE), the kinetic temperature to be
160~K,  and the observed transitions (with the exception of OCS 9--8) to
be optically thin.  The assumption of a single temperature instead of a
temperature profile leads to inaccuracies in our estimates of
abundances. However, given the non-centrally peaked geometry of the hot
molecular gas as detected in our observations and the lack of dust
continuum data, as well as observations of multiple spectral lines
constraining the density and temperature profiles, resolving these
uncertainties is beyond the scope of this paper.  We propose the use of
these abundances in combination with upcoming high angular resolution
complementary observations in order to provide constraints for future
chemical models.

Based on the peak brightness temperatures presented in
Table~\ref{basic_results} we conclude that in all likelihood
\hcthreen, SO, and \methanolap\ $3_1$--$4_0$ transitions are also
optically thick. In the absence of observations of rarer species, we
use the estimated column densities as lower limits and/or guidelines.
For both \methanol\ and \ethylcyanide\ we adopt the largest among the
column densities derived from the different observed transitions. The
\methanol\ $7_2$--$6_3$ A$^{-}$ transition gives the highest column
density and it has a peak brightness temperature of 82~K, which
suggests that it is still optically thin.   The column density of OCS
9--8 is derived by using the OC$^{34}$S column densities and assuming
the relative Galactic abundance of $^{32}S$/$^{34}S$ $\sim 22$.
Table~\ref{colden} presents the observed column densities and
abundances of the different species at the positions of the
north-eastern (NE) and south-eastern (SE) peaks, abundances of NH$_3$
\citep{heaton1989}, \methylcyanide\ \citep{watt1999}, along with
abundances of different species in other ``prototypical" hot cores.

The present dataset does not provide an independent estimate of the
total molecular hydrogen column density, \nhtwo, in the ultracompact
core \citep[UCC;][]{heaton1993} of G34.26+0.15. Estimates of \nhtwo\
in the core, mostly based on a number of single-dish and
interferometric observations of different chemical species, vary by
almost an order of magnitude. Using the peak NH$_3$ abundance of
2$\times$10$^{-5}$ \citep{millar1997}  and the observed NH$_3$ column
density of 7$\times$10$^{18}$~\cmsq\ \citep{heaton1989},
\citet{watt1999} estimated \nhtwo\ to be 4$\times 10^{23}$~\cmsq. In
contrast based on NH$_3$ \citep{heaton1989} and CH$_3$CN
\citep{akeson1996} observations, the \nhtwo\ is estimated to be
3--$7\times 10^{24}$~\cmsq. The lower H$_2$ column densities deduced
by \citet{watt1999} are however consistent with the non-detection of
\CeiO\ at the center of the core.  Following \citet{watt1999} here we
adopt the \nhtwo\ of the UCC within G34.26+0.15 to be
4$\times$10$^{23}$~\cmsq\ in order to calculate the observed
abundances of the observed chemical species relative to molecular
hydrogen (Table~\ref{colden}).  The \methylcyanide\ and \ammonia\
observations that have angular resolution comparable to our observations
justify the adopted value of \nhtwo.

The derived column densities of the complex molecules like \methanol,
\methylformate, \ethylcyanide\ and \dimethylether\ agree reasonably
well with the previous observations (at 13\arcsec$\times$8\arcsec\
resolution) by \citet{mehringer1996}.  Following Table~\ref{colden} we
find that at the two rather clearly segregated peak positions within
the inner regions of the HMC in G34.26+0.15, the abundances of
\dimethylether, \ethylcyanide, \hcthreen, \formamide, and SO$_2$
differ by factors $< 3$.

\section{Hot core chemistry in G34.26+0.15}

We have detected several of the N-, O- and S-bearing parent and
daughter molecules in the G34.26+0.15 hot core region.  Here we
discuss the relative abundances of the different species in the light
of the abundances observed in a few well-known hot core regions
(Table~\ref{colden}). We also compare the abundances relative to a few
well-known chemical models for hot cores, these include: chemical
models for G34.26+0.15 by \citet{millar1997} and \citet{nomura2004},
both at $t_{\rm core}$ = 10$^4$~yr after the grain mantle evaporation;
and  the model for the Orion compact ridge at $t_{\rm core}$ =
3.3$\times10^4$~yr  by \citet{caselli1993}. The model by
\citet{caselli1993} though not fine-tuned for G34.26+0.15, is relevant
because previous chemical study by \citet{mehringer1996} suggested
that the observed abundances can be explained reasonably well by this
model at $t_{\rm core}=10^4$~yr.

The chemical model for G34.26+0.15 presented by \citet{nomura2004} is
essentially an improvement of the model by \citet{millar1997} in terms
of using latest collision rates and chemistry. However for the model
by \citet{nomura2004} column densities are available only for the
entire source, rather than explicitly for the UCC or the inner regions
of the HMC in G34.26+0.15.  Since, the enhanced angular resolutions of
the present BIMA observations enables us to probe the inner regions of
the HMC directly, the column density calculations for the UCC by
\citet{millar1997} are still relevant for our work. Further we note
the column densities of the entire source (core+halo) as estimated by
these two models agree reasonably well.

Figure~\ref{modelcomp} compares the abundances of the different
species relative to the molecular hydrogen abundance as observed at
the NE peak of G34.26+0.15 with abundances predicted by the different
chemical models.  In the following sections we discuss the abundances
of the O-, N- and S-bearing molecular species separately and compare
them with abundances seen in other hot cores. 

\subsection{Oxygenated species} 

We have detected \methanol, \dimethylether\ and \methylformate.
Figure~\ref{modelcomp} shows that there is a huge discrepancy between
the abundances of \methanol\ as predicted by \citet{caselli1993} and
\citet{millar1997}, with the latter model reproducing the observed
abundances in G34.26+0.15 exactly.  All the chemical models predict
the relative abundance \methylformate /\methanol\ to be between
10$^{-3}$ and 10$^{-2}$ and do not reproduce the high abundance ratio
of 0.1 as observed in most hot cores (Table~\ref{colden}). On the
other hand a relative abundance of \dimethylether/\methanol\ $\sim
10^{-3}$ is predicted by all the models, which is more or less
consistent with the observed range of values in the other hot cores.
Most of the chemical models for hot cores, including those considered
in this paper rely solely on gas-phase chemistry to account for the formation
of the large O-bearing complex molecules. However, \citet{horn2004}
have shown that the gas-phase processes are highly inefficient in
producing \methylformate\ and can never reproduce the observed
abundances. Recently \citet{garrod2006} has proposed a slower warm-up
phase of the hot core and a combination of grain-surface and gas-phase
chemistry in the hot cores, which better reproduce the observed
abundances of \methylformate.

\subsection{Nitrogenated Species}

Table~\ref{colden} suggests that the abundances of the N-bearing
species, \ammonia, \methylcyanide, \ethylcyanide, \hcthreen, and
\formamide\ in G34.26+0.15 lie within the range of values seen in
other hot cores.

\begin{figure}
\begin{center}
\includegraphics[angle=0,width=10.0cm]{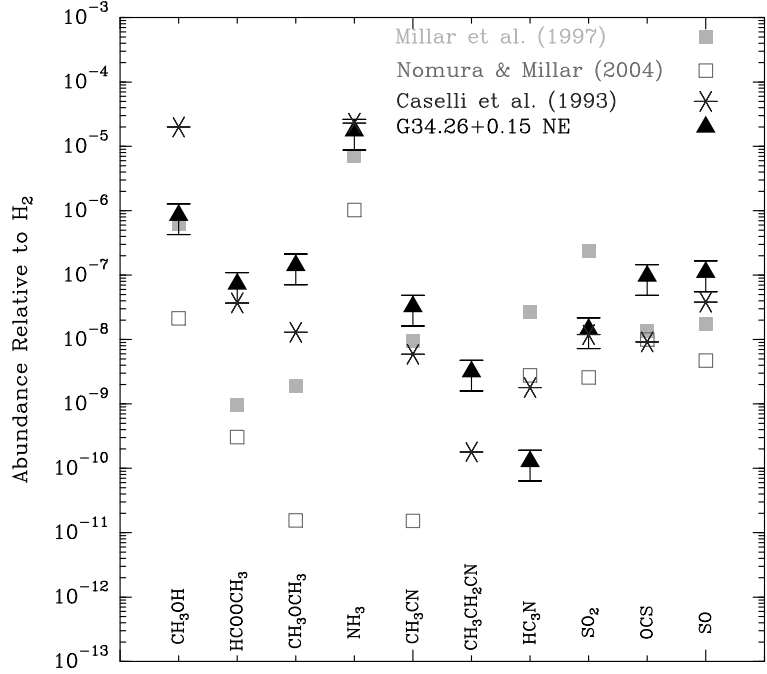}
\caption{Comparison of column densities observed at the north-eastern
(NE) peak of G34.26+0.15 with different chemical models. The observed 
relative abundances are derived by assuming \nhtwo\ =
4$\times10^{23}$~\cmsq. The errorbars correspond to 50\% errors in
abundance estimates. \citet{caselli1993} corresponds to the chemical
model for the Orion Compact Ridge at $t_{\rm core}=3.3\times10^4$~yr.
\citet{millar1997} corresponds to the chemical model for the
ultracompact core, for which  \nhtwo\ = 2.7$\times10^{23}$~\cmsq\ at
$t_{\rm core}=10^4$~yr. \citet{nomura2004} corresponds to a chemical
model for the entire source G34.26+0.15, for a line of sight through
the center of the core at $t_{\rm core}=10^4$~yr, with \nhtwo\ =
3.6~10$^{24}$.
\label{modelcomp}}
\end{center}
\end{figure}

Discovery of \ammonia\ ice has lead to the conclusion that \ammonia\
is indeed a parent species evaporated from the grain mantle.  Presence
or absence of \ammonia\ in the grain mantle plays an important role in
determining the abundances of the different daughter species
\citep{rodgers2001}.  Figure~\ref{modelcomp} suggests that while
\citet{caselli1993} reproduces the \ammonia\ abundance seen in
G34.26+0.15; both \citet{millar1997} and \citet{nomura2004} fall short
by a factor of $\sim 10$ or more, with the latter model being much
worse.

\methylcyanide, a symmetric top molecule, is  likely to form via
radiative association of CH$_3^+$ with HCN or CN with CH$_3$
\citep{charnley1992,millar1997}.  The chemical models for the UCC of
G34.26+0.15 \citep{millar1997} and the Orion Compact Ridge
\citep{caselli1993} predict \methylcyanide\ abundances 30--50 times
lower than the observed value in G34.26, while the core+halo model of
\citet{nomura2004} predicts an even lower \methylcyanide\ abundance
(Fig.~\ref{modelcomp}).

The high abundances of nitrile molecules like \ethylcyanide\ as found
in many hot cores cannot be reproduced by gas phase reactions alone.
Hydrogenation of accreted \hcthreen\ and \methylcyanide\ to produce
\ethylcyanide\ that when evaporates, is proposed to be a more viable
path for \ethylcyanide\ formation in hot cores
\citep{charnley1992,caselli1993}.  The chemical model for Orion
Compact Ridge by \citet{caselli1993} predicts an abundance that is 10
times lower than observed in G34.26. No predictions for the abundance
of \ethylcyanide\ are available from the other two chemical models in
consideration here.

\hcthreen\ is thought of as a cold-cloud tracer but should also be
formed efficiently in the hot gas if C$_2$H$_2$ is evaporated from
grain mantles \citep{millar1997}.  Figure~\ref{modelcomp} shows that
all the chemical models predict an \hcthreen\ abundance larger by more
than a factor of 100 than the value observed in G34.26. This is
consistent with our estimate that \hcthreen\ is optically thick
(Sec.~\ref{kintemp}) in G34.26+0.15.

\formamide\ is most likely formed by atom addition to HCO on grain
mantles and then evaporate \citep{tielens1997}. \citet{bernstein1995}
found \formamide\ to be formed upon UV photolysis and warm-up of
H$_2$O:\methanol:CO:\ammonia = 100:50:10:110 mixture. However, most
of the chemical models so far do not consider either \ethylcyanide\ or
\formamide\ as parent species and there are no predictions of
abundances available so far. 

\subsection{Sulphuretted species}

Based on single-dish observations of G34.26+0.15 and subsequent
chemical modeling, \citet{hatchell1998} inferred that the OCS
emission, which requires rather high excitation energy, can not be
produced in the extended halo; the rotation temperature derived from
SO$_2$ warrants a core origin;  and the SO lines are more likely to
arise from the halo. We estimate both OCS and SO to be optically thick
and to have peak brightness temperatures characteristic of the inner
regions.
Table~\ref{colden} shows that abundance of OCS as observed in
G34.26+0.15 is larger than those observed in many hot cores. The
abundance of SO in G34.26+0.15 is similar to that in Orion, while Sgr
B2(N) and G327.6--0.3 show much lower abundances.  Note that the SO
abundance derived here is likely a lower limit due to the optically
thick SO line. The SO$_2$ abundance in G34.26+0.15, is similar to the
abundance seen in Sgr B2(N) and both are lower by a factor of 10 than
the values seen in Orion and G327.3--0.6.  It was initially pointed out
by \citet{charnley1997} and \citet{hatchell1998} that the relative
abundance ratios of SO, SO$_2$ and H$_2$S could be used to estimate
the age of the hot cores of the massive protostars.  However,
\citet{wakelam2004} has re-considered this issue and conclude that
none of these ratios can be used by itself to estimate the age, since
the ratios depend at least as strongly on the physical conditions and
on the adopted grain mantle composition as on time.

Figure~\ref{modelcomp} shows that all the chemical models under
consideration here predict very similar OCS abundances, although
somewhat less than that observed in G34.26+0.15. The model by
\citet{caselli1993} reproduces the observed SO abundance in
G34.26+0.15  reasonably well, while the other two models predict
abundances lower by factors of 5--10. Since we estimate SO in
G34.26+0.15 to be optically thick, it is hard to reconcile to the
higher abundance derived observationally.  The model by
\citet{caselli1993} reproduces the  observed abundances of SO$_2$ for
G34.26+0.15, while the models by \citet{millar1997} and
\citet{nomura2004} respectively overestimates and underestimates the
observed values.

\subsection{Summary of comparison with chemical models}

None of the chemical models considered here reproduce the abundances
of all the observed species satisfactorily.  These models like all
the other chemical and radiative transfer models for HMCs assume the
energizing source to be at the center, with centrally peaked
temperature and density profiles. In addition, there are reasonable
arguments in favor, as well as, against the 
complex molecule formation mechanisms (grain-mantle evaporation,
grain-surface chemistry, gas-phase chemistry).
Thus, from the physical point of view
all these models are not directly applicable to G34.26+0.15, which as
our observations show is externally heated and additionally might be
harboring multiple unresolved HMCs.

There are two main uncertainties in our estimates of the column
densities of the different species: {\em (i)} we have assumed a single
kinetic temperature,  rather than a temperature profile,  the
value of which may also be off by 50~K for some of the species and
{\em (ii)} the \nhtwo\ estimate is also not accurate.  Both of these
uncertainties can only be addressed with complementary observations at
comparable angular resolutions. The errorbars drawn at the 50\% level
in Fig.~\ref{modelcomp} attempt to quantify these uncertainties in
addition to the statistical errors and the errors in absolute
calibration.  Since the difference in abundances between the NE and SE
peaks of G34.26+0.15 never exceeds a factor of 3 and the discrepancy
between the observed and predicted abundances are of the order of
factors of 10 or more, our conclusions would not have been
significantly different had we compared the observed abundances at the
SE peak with the model predictions.  Further, we have shown that the
abundances observed in G34.26+0.15 are not atypical compared to other
Galactic hot cores.  Thus, discrepancies by factors of 10 to 100
between the abundances observed in G34.26+0.15 and the predictions of
the chemical models can not be explained by the estimated errors in
abundance calculations. Given the inappropriateness of the
existing chemical models and uncertainties in the derived relative
abundances we consider any age/timescale related discussion for the
HMC in G34.26+0.15 beyond the scope of this paper.

\section{Discussion
\label{discussion}}

Despite the present consensus on the contemporaneous existence of
\ammonia\ and \methanol\ bearing grain mantles, chemical differentiation
in hot core regions is also well-established from observations.  Such
chemical differentiation was prominently noticed in the two clumps in
the Orion-KL cloud core, very close to the luminous and massive young
stellar object, IRc2: the Hot core and the Compact Ridge. While the
warmer and denser Hot Core shows unusually high abundances of H-rich
complex N-bearing molecules like CH$_2$CHCN and \ethylcyanide, the
Compact Ridge is characterized by high abundances of large
oxygen-bearing molecules like \methanol, \methylformate\ and
\dimethylether. However, later observations of the Orion-KL hot cores by
\citet{sutton1995} suggest that the chemical differentiation is probably
not as drastic as it was first thought to be.  Similar chemical
differentiation is observed between the two hot cores W3(H$_2$O) and
W3(OH) in the W3 region \citep{wyrowski1997,wyrowski1999}.  These
sources are spatially offset by about 0.06 pc and yet they exhibit clear
signs of N/O differentiation. Both sources show emission from \methanol,
H$_2$CO, \dimethylether,  and \methylformate, but only the maser source
shows significant emission from \ethylcyanide, \hcthreen, and SO2
\citep{wyrowski1997,wyrowski1999}, as well as from HCN
\citep{turner1984} and \methylcyanide\ \citep{wilson1993}.

Several chemical models have partially explained the chemistry of both
the Hot Core \& the Compact Ridge \citep{caselli1993,charnley1992} in
Orion.  \citet{rodgers2001} have shown that the evaporation of \ammonia\
rich grain mantles inhibit high abundances of O-bearing large molecules,
though the already injected alcohols remain abundant for much longer.
This leads to the formation of N-rich hot cores. \citet{rodgers2003}
further showed that in collapsing cores both N-rich and O-rich species
tend to co-exist.  However, it might be possible to derive the age of
the cores based on the relative abundance of the two types of species
because \dimethylether\ is found to be more abundant at earlier times
whereas HCN and \methylcyanide\ form later on \citep{rodgers2003}.
Observational detection of large number of daughter species in hot cores
suggest that the chemical timescale needs to be shorter than the
dynamical timescale, {\em i.e.,} the cores need to be gravitationally
supported up to 10$^4$~yrs \citep{rodgers2003}.

Within the inner regions of G34.26+0.15 we find that the N-bearing
species tend to peak to the south-east, while the peaks of the
O-bearing species are concentrated to the north-east of the continuum
peak. The NE and the SE peaks are primarily separated along the
north-south direction by 0\farcs8, which at a distance of 3.7~kpc
translates to 0.014~pc.  For comparison, the Hot Core and the Compact
Ridge regions in the Orion-KL cloud  at a distance of 450~pc are
separated by 0.018~pc.   Thus, given the larger distance to the
source and the angular resolution of the present observations, the
possibility of the two emission peaks actually belonging to two
different cores, one being N-rich and the other being O-rich, can not
be ruled out. Although the two peak positions in G34.26 do not show
large difference in \methanol\ abundances, differences in the
abundances of \dimethylether, \ethylcyanide, \hcthreen\ and
\formamide\ show contrasts quite similar to what is seen in Orion-KL
\citep{sutton1995}. {Additional high angular resolution observations
of molecular lines are required to derive a better estimate of the
temperature distribution, in order to improve the derived abundances
at the two peak positions seen within the inner regions of the HMC in
G34.26+0.15}.

\section{Summary}

We have presented high angular resolution mapping observations of the
HMC in G34.26+0.15 tracing different chemical species characteristic
of hot cores. The higher angular resolution enables us to probe the
inner regions of the hot molecular gas. The observations presented
here is only a first step and needs to be augmented by observations of
standard temperature and density sensitive molecular transitions to
constrain the physical attributes of the region which can then be used
as constraints for future chemical models.

We do not detect any evidence for an energizing source at the center
of the hot core, the continuum peak at 2.8~mm is consistent with the
free-free radio continuum emission from component C of the UC \HII\
region.  The intensity distribution of the various molecular tracers
do not peak at the same position, the peaks of none of the
distributions is located either  at the position of the continuum peak
or the positions of the H$_2$O and OH masers. The temperature and
density distributions of the inner regions of the hot cores can not be
determined from the present observations. Based on intensity and
velocity distribution, only component C, the most evolved of the \HII\
regions associated with G34.26+0.15 appears to influence the
energetics of the hot molecular gas. The kinematics of the hot core as
derived from the velocity distribution of the molecular tracers is not
strongly influenced by the \HII\ region dynamics.

Within the inner regions of the hot core in G34.26+0.15 we find that
the nitrogen and oxygen-bearing species tend to show a dichotomy and
peak at different positions, separated by 0\farcs8, the spatial
separation being similar to the two hot/warm cores identified in the
Orion.  We propose that as in Orion and in W3(OH), (i) these peak
positions may indeed be separate regions of chemical enrichment,
resolution of which would require even higher angular resolution
observation and (ii) may arise due to the external influence of the
neighboring \HII\ regions.  We have estimated the abundances of the
observed molecular species at the two peak positions assuming a single
kinetic temperature of 160~K under conditions of LTE.

The high angular resolution observations presented here, provides
overwhelming evidence in favour of the hot molecular gas at spatial
scales of 0.018~pc being externally heated. This together with the
clumpiness of the region and possible existence of multiple HMCs in
the region, implies geometries more complicated than those considered
by the state-of-art chemical models. For the sake of completeness, we
follow \citet{wyrowski1999} to derive a crude upper limit for the
luminosity of the internal energizing source if any, using the kinetic
temperature derived in this paper and Stefan-Boltzmann law.  The
luminosity of the hypothetical and so far undetected internal source
is estimated to be 8.9$\times$10$^4$~\lsun\ and this corresponds to a
O7.5 ZAMS star \citep{panagia1973}.

Comparison of the abundances observed in G34.26+0.15 with currently
available somewhat inappropriate chemical models for hot cores by
\citet{caselli1993}, \citet{millar1997}, and \citet{nomura2004}
suggest that the nitrogen richness of the region can only be explained
by the evaporation of ammonia-rich ice-mantles. However, this implies
that the abundances of the oxygenated daughter species be suppressed
in the presence of ammonia, which is not the case in G34.26+0.15. This
can mean any of the following (i) the oxygenated species in question
are also formed on grain mantles, (ii) the two regions are actually
separate from each other or (iii) there may be additional gas-phase
processes which can progress efficiently even in the presence of
ammonia to create high abundances of oxygenated daughter species,
which the chemical models do not yet consider.  Comparison of
abundances observed in a few hot cores (in addition to G34.26+0.15)
and the chemical models show that the chemical models do not yet
consistently explain the abundances of all species in any of the hot
cores.  The complicated geometry of G34.26+0.15 makes direct
comparison of the observed abundances with the chemical models even
more difficult.

One of the major caveats of the estimates of the abundances presented
here in the hot molecular gas is the assumption of a single kinetic
temperature characterizing the emission of the different species. This
is a consequence of the lack of complementary observations of
molecular species at high angular resolutions to study the temperature
and physical structure of the hot molecular gas. In order to
understand the true nature of the different peak positions in
G34.26+0.15 as identified in the BIMA data and to provide proper
physical constraints for the chemical models for the region,
sub-arcsecond resolution observations are necessary. 

\acknowledgments

This material is based upon work supported by the National Science
Foundation under Grant No. AST-0228974.  This research has made use of
NASA's Astrophysics Data System. LWL acknowledges support from the
Laboratory for Astronomical Imaging at the University of Illinois and
NSF under grant AST-0228953.

\clearpage

\begin{table}
\begin{center}
\caption{Details of BIMA observations of G34.26+0.15
\label{obsdetails}}
\begin{tabular}{cccc}
\tableline
\tableline
Frequency & Configuration & Date & T$_{\rm sys}$ \\
GHz & & & K\\
\tableline
87 & A & 23 Dec 1999 & 180\\
 & B & 26 Feb 2000 & 450\\
107 & A & 29 Jan 2000 & 200\\
 & B & 10 Mar 2000 & 350\\
109 & A & 6 Feb 2000 & 250\\
 & B & 13 Mar 2000 & 300\\
\tableline
\tableline
\end{tabular}
\end{center}
\end{table}

\clearpage

\begin{table}
\begin{center}
\caption{Spectroscopic parameters of molecular transitions detected by 
1\arcsec\ resolution BIMA observations of G34.26+0.15}
\label{speclines}
\vspace*{0.3cm}
\begin{tabular}{llllll}
\tableline
\tableline
Species & Transition & Frequency & E$_u/k$T  & $S_{ij}$ \\ 
&  & GHz & K &  & \\ 
\tableline
$\rm CH_3OH$ &  $7_2$--$6_3$~A$^-$                 & 86.6155080  &
 102.7 & 0.644 \\ 
 &  $7_2$--$6_3$~A$^+$                & 86.9030180  &
 102.7 & 0.644 \\ 
&  $3_1$--$4_0$~A$^+$                 & 107.0138500 &
28.36 & 1.43   \\ 
&  $15_{-2}$--$15_1$~E2               & 107.15992   & 
305 & 1.24 \\ 

$\rm HCOOCH_3$ &  $8_{2,6}$--$7_{2,5}$~E           & 103.46659   &
 24.66 & 19.992 \\ 
$\rm CH_3OCH_3$& $13_{1,12}$--$13_{0,13}$~ EA+AE & 105.7683438 &
86.02 & 5.405 \\ 
$\rm NH_2CHO$& $5_{2,4}$--$4_{2,3}$              & 105.972601  &
27.2 & 4.2 \\ 
$\rm OC^{34}S$& 9--8                             & 106.7873889 &
25.7 & 9.0 \\ 
$\rm O^{13}CS$& 9-8                              & 109.1108477 &
 26.2 & 9.0 \\ 
OCS &9--8                                        & 109.4630630 &
26.28 & 9.0 \\ 
$\rm CH_3CH_2CN$& $12_{2,11}$--$11_{2,10}$       & 107.0435270 &
37.91 & 11.662 \\ 
& $12_{1,11}$--$11_{1,10}$       & 109.6502630 &
 35.42 & 11.906 \\ 
$\rm SO_2$& $27_{3,25}$--$26_{4,22}$             & 107.0602085 &
 369.5 & 3.102 \\ 
$\rm HC_3N$& 12--11                              & 109.1736340 &
 34.07 & 12.0 \\ 
SO& $3_2$--$2_1$                                 & 109.2522200 &
 21.05 & 1.510 \\ 
\tableline
\end{tabular}
\end{center}

\tablecomments{The frequency and energy of $\rm CH_3OH$
$15_{-2}$--$15_{1}$~E2 taken from Sutton et al. (2004). All other data
frequencies and energies were taken from the JPL Molecular
Spectroscopy Catalog (http://spec.jpl.nasa.gov/).  The $\rm CH_3OCH_3$
line is assumed to be in the AE state.}
\end{table}

\clearpage

\begin{table}
\begin{center}
\caption{Basic observational results of the chemical survey towards 
G34.26+0.15. Synthesized beam sizes at 87~GHz, 107~GHz and 109~GHz are
1\farcs48$\times$1\farcs25, 1\farcs02$\times$0\farcs85 and
0\farcs99$\times$0\farcs89 respectively.
\label{basic_results}}
\begin{tabular}{lccccccc}
\tableline
\tableline
Line & Beam & $F_{\rm peak}$ & $T_{\rm B,peak}$ & $v_{\rm cen}$ & $v_{\rm fwhm}$ &  RMS\\
     &  (arcsec) &  (Jy~beam$^{-1}$) &   (K)               &  (\kms)
     & (\kms) &  (mJy~beam$^{-1}$)\\
\tableline
H53$\beta$                                      
&1\farcs48$\times$1\farcs25 & 0.454  & 42  & $56.0\pm1.8$   &
$45.2\pm4.7$  & 66.3 \\
$\rm CH_3OH$ $7_2$--$6_3$~A$^-$                 
& 1\farcs48$\times$1\farcs25 & 0.934  & 82  & $59.1\pm0.3$  &
$6.2\pm1.0$  & 69.1\\
$\rm CH_3OH$  $7_2$--$6_3$~A$^+$                
& 1\farcs48$\times$1\farcs25 & 0.952  & 83  & $59.3\pm0.3$  &
$7.0\pm0.5$  & 69.6\\
$\rm HCOOCH_3$ $8_{2,6}$--$7_{2,5}$~E
& 1\farcs02$\times$0\farcs85& 0.240  & 31  & $56.3\pm2.0$  &
$4.7\pm1.0$  & 35.6\\
$\rm CH_3OCH_3$ $13_{1,12}$--$13_{0,13}$ &
1\farcs02$\times$0\farcs85 & 0.218  & 26 &  49.3\tablenotemark{a} & \ldots & 55.1\\
$\rm NH_2CHO$ $5_{2,4}$--$4_{2,3}$              &
1\farcs02$\times$0\farcs85 & 0.383  & 46  & $55.9\pm0.3$  &
$7.3\pm07$  & 106.\\
$\rm OC^{34}S$ 9--8                             &
1\farcs02$\times$0\farcs85 & 0.394  & 50  & $56.0\pm0.3$  &
$4.0\pm0.7$  & 86.1\\
$\rm CH_3OH$ $3_1$--$4_0$~A$^+$                 &
1\farcs02$\times$0\farcs85 & 1.48   & 187 & $58.4\pm0.2$  &
$6.0\pm0.3$  & 44.9\\
$\rm CH_3CH_2CN$ $12_{2,11}$--$11_{2,10}$       &
1\farcs02$\times$0\farcs85 & 0.365  & 46  & $55.9\pm0.1$  &
$4.4\pm0.5$  & 45.3\\
$\rm SO_2$ $27_{3,25}$--$26_{4,22}$             &
1\farcs02$\times$0\farcs85 & 0.198  & 25  & $56.6\pm1.0$  &
$6.3\pm1.5$  & 45.3\\
$\rm CH_3OH$ $15_{-2}$--$15_1$~E2               &
1\farcs02$\times$0\farcs85 & 0.643  & 81  & $60.0\pm0.1$  &
$6.3\pm0.3$  & 94.9\\
$\rm O^{13}CS$ 9-8                              &
0\farcs99$\times$0\farcs89 & 0.165  & 20  & 55.6\tablenotemark{a}   &  \ldots& 67.5\\
$\rm HC_3N$ 12--11                              &
0\farcs99$\times$0\farcs89 & 1.23   & 143 & $56.5\pm0.1$  &
$6.4\pm0.1$  & 64.3\\
SO $3_2$--$2_1$                                 &
0\farcs99$\times$0\farcs89 & 1.27   & 147 & $58.0\pm0.1$  &
$4.6\pm0.3$  & 123.1\\
OCS 9--8                                        &
0\farcs99$\times$0\farcs89 & 1.44   & 167 & $58.4\pm0.2$  &
$5.2\pm0.4$  & 127.0\\
$\rm CH_3CH_2CN$ $12_{1,11}$--$11_{1,10}$       &
0\farcs99$\times$0\farcs89 & 0.360  & 42  & $58.5\pm0.6$  &
$5.0\pm1.3$  & 65.2\\
\tableline
\tableline
\end{tabular}
\end{center}

\tablenotetext{a}{No Gaussian fit was possible to the spectrum; $v_{\rm cen}$
corresponds to the velocity at which the peak intensity is observed.}
\end{table}

\clearpage

\begin{deluxetable}{lllllllll}
\tablecaption{Chemical abundances of the species observed towards
G34.26+0.15. Assumed $T_{\rm kin}$ = 160~K. $N$ and $X$ refer to
column densities and abundances relative to molecular hydrogen
respectively. NE and SE refer to the north-eastern and south-eastern
peaks respectively. Column densities of optically thick lines of OCS
9--8 has been derived from OC$^{34}$S assuming $^{32}$S/$^{34}$S = 22.
The relative abundances for the compact core have been calculated
using \nhtwo = 4$\times 10^{23}$~\cmsq\ \citep{watt1999}. $a(b)$
refers to $a\times10^b$. Considering the errors in absolute
calibration, statistical errors, errors due to the assumption
of a single $T_{\rm kin}$, and the error in \nhtwo\ we conservatively 
estimate an error of 50\% in the calculated abundances.
\label{colden}}
\tabletypesize{\footnotesize}
\tablecolumns{10}
\tablehead{\colhead{Species} & \colhead{$N_{\rm NE}$} &
\colhead{$X_{\rm NE}$} & \colhead{$N_{\rm SE}$} & \colhead{$X_{\rm SE}$} &
\colhead{OMC\tablenotemark{a}} &  \colhead{OMC\tablenotemark{a} }& \colhead{SgrB2
(N)\tablenotemark{b}} &
\colhead{G327.3--0.6\tablenotemark{c}} \\
& \colhead{\cmsq} & &\colhead{\cmsq} &  &\colhead{Hot Core}
& \colhead{Compact Ridge}&&} 
\startdata
\methanol &   3.4(17) &   8.5(\phn-7) &  2.6(17) &   6.4(\phn-7) &  
1.4(\phn-7) & 4.0(\phn-7)   & 2.0(\phn-7)    & 2.0(\phn-5)   \\
\methylformate &   2.9(16) &   7.3(\phn-8) &  2.7(16) & 6.8(\phn-8) & 
1.4(\phn-8) & 3.0(\phn-8) & 1.0(\phn-9)         & 2.0(\phn-6) \\
\dimethylether &   5.7(16) &   1.4(\phn-7) &  3.4(16) &   8.5(\phn-8) & 
8.0(\phn-9) & 1.9(\phn-8) & 3.0(\phn-9)    & 3.4(\phn-7) \\
\ammonia\tablenotemark{d}& 7.0(18) & 1.8(-5) &\ldots&\ldots & 
6.7(\phn-8) & 5.7(\phn-7) & \ldots         & \ldots       \\
\methylcyanide\tablenotemark{d} & 1.3(16) & 3.3(\phn-8) & \ldots & \ldots  & 
4.0(\phn-9) & 5.0(\phn-9) & 3.0(\phn-8)    & 7.0(\phn-7) \\
\ethylcyanide &   1.3(15) &   3.2(\phn-9) &  2.7(15) & 6.8(\phn-9) &  
3.0(\phn-9)   & 5.0(\phn-9)   & 6.0(-10)    & 4.0(\phn-7)  \\
\hcthreen &   5.1(13) &   1.3(-10) &  1.3(14) &   3.3(-10) & 
1.8(\phn-9) & 6.0(\phn-9) & 5.0(\phn-9)    & 3.0(-11)   \\
NH$_2$CHO &   \ldots &   \ldots &  2.5(15) &   6.2(\phn-9) & 
1.4(-10)    & 3.0(-10)    & 2.0(-10)       & 2.0(\phn-8)    \\
SO$_2$ &   5.8(15) &   1.4(\phn-8) &  1.5(16) &   3.7(\phn-8) & 
1.2(\phn-7) & 1.6(\phn-7) & 3.0(\phn-8)      & 2.0(\phn-7)    \\
OC$^{34}$S &   1.8(15) &   4.4(\phn-9) &  2.2(15) &   5.6(\phn-9) & 
\ldots      & \ldots      & 3.0(-10)         & 1.0(\phn-9) \\
O$^{13}$CS &   2.0(15) &   5.1(\phn-9) &  1.4(15) &   3.5(\phn-9) &
\ldots      & \ldots      & 7.0(-10)       & 7.0(-10)    \\
OCS &   3.9(16) &   9.7(\phn-8) &  4.9(16) &   1.2(\phn-7) & 
1.1(\phn-8) & 3.0(\phn-8) & $>$2.0(\phn-9) & $>$2(\phn-9)\\
SO &   4.4(16) &   1.1(\phn-7) &  5.3(16) &   1.3(\phn-7) &
1.9(\phn-7) & 3.0(\phn-7) & 2.0(\phn-8)    & $>$3(\phn-9) \\
\enddata
\tablenotetext{a}{\citet{sutton1995,caselli1993}}
\tablenotetext{b}{\citet{nummelin2000,liu1999}}
\tablenotetext{c}{\citet{gibb2000}}
\tablenotetext{d}{Column densities for the entire hot core from
\citet{watt1999} (\methylcyanide) and \citet{heaton1989} (\ammonia)}
\end{deluxetable}

\clearpage

\clearpage

\clearpage

\clearpage

\clearpage

\begin{thebibliography}{}


\bibitem[Akeson \& Carlstrom(1996)]{akeson1996} Akeson, R.~L., \& 
Carlstrom, J.~E.\ 1996, \apj, 470, 528
\bibitem[Avalos et al.(2006)]{avalos2006} Avalos, M., Lizano, S., 
Rodr{\'{\i}}guez, L.~F., Franco-Hern{\'a}ndez, R., \& Moran, J.~M.\ 2006, 
\apj, 641, 406
\bibitem[Beltr{\'a}n et al.(2005)]{beltran2005} Beltr{\'a}n, M.~T., 
Cesaroni, R., Neri, R., Codella, C., Furuya, R.~S., Testi, L., \& Olmi, L.\ 
2005, \aap, 435, 901 
\bibitem[Beltr{\'a}n et al.(2004)]{beltran2004} Beltr{\'a}n, M.~T., 
Cesaroni, R., Neri, R., Codella, C., Furuya, R.~S., Testi, L., \& Olmi, L.\ 
2004, \apjl, 601, L187
\bibitem[Bernstein et al.(1995)]{bernstein1995} Bernstein, M.~P., 
Sandford, S.~A., Allamandola, L.~J., Chang, S., \& Scharberg, M.~A.\ 1995, 
\apj, 454, 327
\bibitem[Campbell et al.(2000)]{campbell2000} Campbell, M.~F., 
Garland, C.~A., Deutsch, L.~K., Hora, J.~L., Fazio, G.~G., Dayal, A., \& 
Hoffmann, W.~F.\ 2000, \apj, 536, 816 
\bibitem[Campbell et al.(2004)]{campbell2004} Campbell, M.~F., 
Harvey, P.~M., Lester, D.~F., \& Clark, D.~M.\ 2004, \apj, 600, 254 
\bibitem[Carral \& Welch(1992)]{carral1992} Carral, P., \& Welch, 
W.~J.\ 1992, \apj, 385, 244
\bibitem[Caselli et al.(1993)]{caselli1993} Caselli, P., Hasegawa, 
T.~I., \& Herbst, E.\ 1993, \apj, 408, 548
\bibitem[Cesaroni et al.(1998)]{cesaroni1998} Cesaroni, R., Hofner, 
P., Walmsley, C.~M., \& Churchwell, E.\ 1998, \aap, 331, 709
\bibitem[Cesaroni(2005)]{cesaroni2005} Cesaroni, R.\ 2005, IAU 
Symposium, 227, 59
\bibitem[Charnley et al.(1992)]{charnley1992} Charnley, S.~B., 
Tielens, A.~G.~G.~M., \& Millar, T.~J.\ 1992, \apjl, 399, L71
\bibitem[Charnley(1997)]{charnley1997} Charnley, S.~B.\ 1997, \apj, 
481, 396
\bibitem[Forster \& Caswell(1989)]{forster1989} Forster, J.~R., \& 
Caswell, J.~L.\ 1989, \aap, 213, 339
\bibitem[De Buizer et al.(2002)]{debuizer2002} De Buizer, J.~M., 
Watson, A.~M., Radomski, J.~T., Pi{\~n}a, R.~K., \& Telesco, C.~M.\ 2002, 
\apjl, 564, L101
\bibitem[De Buizer et al.(2003)]{debuizer2003} De Buizer, J.~M., 
Radomski, J.~T., Telesco, C.~M., \& Pi{\~n}a, R.~K.\ 2003, \apj, 598, 1127 
\bibitem[Doty et al.(2006)]{doty2006} Doty, S.~D., van Dishoeck, 
E.~F., \& Tan, J.~C.\ 2006, \aap, 454, L5
\bibitem[Garay et al.(1985)]{garay1985} Garay, G., Reid, M.~J., 
\& Moran, J.~M.\ 1985, \apj, 289, 681 
\bibitem[Garay et al.(1986)]{garay1986} Garay, G., Rodriguez, 
L.~F., \& van Gorkom, J.~H.\ 1986, \apj, 309, 553
\bibitem[Garrod \& Herbst(2006)]{garrod2006} Garrod, R.~T., \& 
Herbst, E.\ 2006, \aap, 457, 927
\bibitem[Gaume et al.(1994)]{gaume1994} Gaume, R.~A., Fey, A.~L., 
\& Claussen, M.~J.\ 1994, \apj, 432, 648
\bibitem[Gibb et al.(2000)]{gibb2000} Gibb, E., Nummelin, A., 
Irvine, W.~M., Whittet, D.~C.~B., \& Bergman, P.\ 2000, \apj, 545, 309
\bibitem[Hatchell et al.(1998)]{hatchell1998} Hatchell, J., 
Thompson, M.~A., Millar, T.~J., \& MacDonald, G.~H.\ 1998, \aap, 338, 713
\bibitem[Hatchell et al.(2001)]{hatchell2001} Hatchell, J., Fuller, 
G.~A., \& Millar, T.~J.\ 2001, \aap, 372, 281
\bibitem[Hatchell \& van der Tak(2003)]{hatchell2003} Hatchell, J., 
\& van der Tak, F.~F.~S.\ 2003, \aap, 409, 589
\bibitem[Heaton et al.(1989)]{heaton1989} Heaton, B.~D., Little, 
L.~T., \& Bishop, I.~S.\ 1989, \aap, 213, 148
\bibitem[Heaton et al.(1993)]{heaton1993} Heaton, B.~D., Little, 
L.~T., Yamashita, T., Davies, S.~R., Cunningham, C.~T., \& Monteiro, T.~S.\ 
1993, \aap, 278, 238
\bibitem[Henkel et al.(1987)]{henkel1987} Henkel, C., Wilson, 
T.~L., \& Mauersberger, R.\ 1987, \aap, 182, 137
\bibitem[Horn et al.(2004)]{horn2004} Horn, A., M{\o}llendal, 
H., Sekiguchi, O., Uggerud, E., Roberts, H., Herbst, E., Viggiano, A.~A., 
\& Fridgen, T.~D.\ 2004, \apj, 611, 605
\bibitem[Hunter et al.(1998)]{hunter1998} Hunter, T.~R., 
Neugebauer, G., Benford, D.~J., Matthews, K., Lis, D.~C., Serabyn, E., \& 
Phillips, T.~G.\ 1998, \apjl, 493, L97
\bibitem[Kim et al.(2000)]{kim2000} Kim, H.-D., Cho, S.-H., 
Chung, H.-S., et al.,
 2000, \apjs, 131, 483
\bibitem[Kuchar \& Bania(1994)]{kuchar1994} Kuchar, T.~A., \& 
Bania, T.~M.\ 1994, \apj, 436, 117
\bibitem[Kurtz et al.(2000)]{kurtz2000} Kurtz, S., Cesaroni, R., 
Churchwell, E., Hofner, P., \& Walmsley, C.~M.\ 2000, Protostars and 
Planets IV, 299
\bibitem[Liu \& Snyder(1999)]{liu1999} Liu, S.-Y., \& Snyder, 
L.~E.\ 1999, \apj, 523, 683 
\bibitem[Lovas et al.(1979)]{lovas1979} Lovas, F.~J., Johnson, 
D.~R., \& Snyder, L.~E.\ 1979, \apjs, 41, 451
\bibitem[MacDonald et al.(1995)]{macdonald1995} MacDonald, G.~H., 
Habing, R.~J., \& Millar, T.~J.\ 1995, \apss, 224, 177
\bibitem[MacDonald et al.(1996)]{macdonald1996} MacDonald, G.~H., 
Gibb, A.~G., Habing, R.~J., \& Millar, T.~J.\ 1996, \aaps, 119, 333
\bibitem[Mehringer \& Snyder(1996)]{mehringer1996} Mehringer, D.~M., 
\& Snyder, L.~E.\ 1996, \apj, 471, 897
\bibitem[Millar(1993)]{millar1993} Millar, T.~J.\ 1993, Dust and 
Chemistry in Astronomy, 249 
\bibitem[Millar et al.(1995)]{millar1995} Millar, T.~J., 
MacDonald, G.~H., \& Habing, R.~J.\ 1995, \mnras, 273, 25
\bibitem[Millar et al.(1997)]{millar1997} Millar, T.~J., 
MacDonald, G.~H., \& Gibb, A.~G.\ 1997, \aap, 325, 1163
\bibitem[Nomura \& Millar(2004)]{nomura2004} Nomura, H., \& 
Millar, T.~J.\ 2004, \aap, 414, 409
\bibitem[Nummelin et al.(2000)]{nummelin2000} Nummelin, A., Bergman, 
P., Hjalmarson, {\AA}., Friberg, P., Irvine, W.~M., Millar, T.~J., Ohishi, 
M., \& Saito, S.\ 2000, \apjs, 128, 213
\bibitem[Panagia(1973)]{panagia1973} Panagia, N.\ 1973, \aj, 78, 
929 
\bibitem[Reid \& Ho(1985)]{reid1985} Reid, M.~J., \& Ho, 
P.~T.~P.\ 1985, \apjl, 288, L17
\bibitem[Rodgers \& Charnley(2001)]{rodgers2001} Rodgers, S.~D., \& 
Charnley, S.~B.\ 2001, \apj, 546, 324 
\bibitem[Rodgers \& Charnley(2003)]{rodgers2003} Rodgers, S.~D., \& 
Charnley, S.~B.\ 2003, \apj, 585, 355
\bibitem[Sault et al.(1995)]{sault1995} Sault, R.~J., Teuben,
P.~J., \& Wright, M.~C.~H.\ 1995, ASP Conf.~Ser.~ 77: Astronomical Data
Analysis Software and Systems IV, 77, 433
\bibitem[Sewilo et al.(2004)]{sewilo2004} Sewilo, M., Churchwell, 
E., Kurtz, S., Goss, W.~M., \& Hofner, P.\ 2004, \apj, 605, 285 
\bibitem[Sutton et al.(1995)]{sutton1995} Sutton, E.~C., Peng, R., 
Danchi, W.~C., Jaminet, P.~A., Sandell, G., \& Russell, A.~P.~G.\ 1995, 
\apjs, 97, 455
\bibitem[Tielens \& Charnley(1997)]{tielens1997} Tielens, 
A.~G.~G.~M., \& Charnley, S.~B.\ 1997, Origins of Life and Evolution of the 
Biosphere, 27, 23
\bibitem[Turner et al.(1974)]{turner1974} Turner, B.~E., Balick, 
B., Cudaback, D.~D., Heiles, C., \& Boyle, R.~J.\ 1974, \apj, 194, 279
\bibitem[Turner \& Welch(1984)]{turner1984} Turner, J.~L., \& 
Welch, W.~J.\ 1984, \apjl, 287, L81
\bibitem[van Buren et al.(1990)]{vanburen1990} van Buren, D., Mac 
Low, M.-M., Wood, D.~O.~S., \& Churchwell, E.\ 1990, \apj, 353, 570
\bibitem[van der Tak(2005)]{vandertak2005} van der Tak, F.~F.~S.\ 
2005.\ The chemistry of high-mass star formation.\ IAU Symposium 227, 
70-79.
\bibitem[van Dishoeck \& Blake(1998)]{vandishoeck1998} van Dishoeck, 
E.~F., \& Blake, G.~A.\ 1998, \araa, 36, 317
\bibitem[Wakelam et al.(2004)]{wakelam2004} Wakelam, V., Caselli, 
P., Ceccarelli, C., Herbst, E., \& Castets, A.\ 2004, \aap, 422, 159
\bibitem[Watt \& Mundy(1999)]{watt1999} Watt, S., \& Mundy, 
L.~G.\ 1999, \apjs, 125, 143
\bibitem[Wilson et al.(1993)]{wilson1993} Wilson, T.~L., Gaume, 
R.~A., \& Johnston, K.~J.\ 1993, \apj, 402, 230
\bibitem[Wilson \& Rood(1994)]{wilson1994} Wilson, T.~L., \& Rood, 
R.\ 1994, \araa, 32, 191
\bibitem[Wood \& Churchwell(1989)]{wood1989} Wood, D.~O.~S., \& 
Churchwell, E.\ 1989, \apjs, 69, 831
\bibitem[Wyrowski et al.(1999)]{wyrowski1999} Wyrowski, F., Schilke, 
P., Walmsley, C.~M., \& Menten, K.~M.\ 1999, \apjl, 514, L43 
\bibitem[Wyrowski et al.(1997)]{wyrowski1997} Wyrowski, F., Hofner, 
P., Schilke, P., Walmsley, C.~M., Wilner, D.~J., \& Wink, J.~E.\ 1997, 
\aap, 320, L17 

\end{thebibliography}
\end{document}